# Olivine-Carbonate Mineralogy of Jezero Crater

A. J. Brown[1], C. E. Viviano[2], and T. A. Goudge[3]

[1]Plancius Research, 1106 Bellevista Ct, Severna Park, MD 21146. [2]Johns Hopkins Applied Physics Laboratory, MD. [3]Jackson School of Geosciences, University of Texas, TX.

Corresponding author: Adrian Brown (adrian.j.brown@nasa.gov)

**Key Points:**

- We describe a correlation between carbonates of Jezero crater and Nili Fossae and the accompanying olivine lithology
- We identify three olivine lithologies and determine variation in Mg-Fe chemistry. Carbonates occur in relatively Fe-rich olivine-lithologies
- We show a formation model for Jezero that accounts for the observations. We discuss how formation mechanisms address the new observations

**Abstract**

A well-preserved, ancient delta deposit, in combination with ample exposures of Mg-carbonate rich materials, make Jezero Crater in Nili Fossae a compelling astrobiological site and a top candidate for future landed missions to Mars. We use CRISM observations to characterize the surface mineralogy of the crater and surrounding watershed. We have identified a three-endmember sequence of olivine-bearing lithologies that we hypothesize are distinguished by their Mg content. We find that Mg-carbonates are consistently identified in association with one of the olivine-bearing lithologies, although that lithology is not fully carbonatized. Surprisingly, this lithology contains relatively Fe-rich olivine. We address a range of formation scenarios, including the possibility that these olivine and carbonate associations are indicators of serpentinization on early Mars. These deposits provide an opportunity for deepening our understanding of early Mars by revealing the thermal history of the martian interior and potentially changes in its tectonic regime with time.

**1 Introduction**

The Nili Fossae region contains a number of proposed landing sites for the upcoming Mars 2020 Rover (Mustard et al., 2013; Williford, 2018). A distinguishing feature of many of these sites is the access to large exposures of carbonate-bearing deposits (Ehlmann et al., 2008b, 2009; Niles et al., 2012; Salvatore et al., 2017; Wray et al., 2016). The origin of these carbonate deposits has been the subject of some discussion in the scientific community (Niles et al., 2012). This results of this study further constrain the conditions of carbonate formation (or carbonization) at Nili Fossae and thus are directly relevant to this ongoing debate. Along with the variety of other carbonate emplacement methods we discuss further below, serpentinization is an especially promising pathway that can occur over a range of geophysical conditions, including in the subsurface at high $p$CO$_2$ (Brown et al., 2010; McSween et al., 2014; Viviano et al., 2013) or near the surface, at low temperature and pressures (Brown et al., 2010; Ehlmann et al., 2009).

Amongst the sites of interest in the Nili Fossae region is the paleolake basin contained within the ~45 km diameter Jezero impact crater (Fassett and Head, 2005). The Jezero crater paleolake is classified as hydrologically open, and was fed by two inlet valleys to the north and west, and drained by an outlet valley to the east (Fassett and Head, 2005). Buffered crater counts of inflowing valley networks indicate that this system ceased fluvial activity by approximately the Noachian-Hesperian boundary (Fassett and Head, 2008), similar to the timing of other large valley network systems on Mars (Fassett and Head, 2008; Hoke and Hynek, 2009). Jezero crater contains two well-exposed fluvio-lacustrine delta deposits (Ehlmann et al., 2008a; Fassett and Head, 2005; Goudge et al., 2015, 2017; Schon et al., 2012) as well as large exposures of both phyllosilicate minerals and carbonates (Ehlmann et al., 2008b, 2008a, 2009, Goudge et al., 2015, 2017). These outcrops make Jezero crater highly appealing as a potential landing site for the Mars 2020 rover, as *in situ* exploration of this region could provide novel insights into both the fluvial sedimentary record and aqueous alteration history of early Mars.

The geologic units within Jezero crater provide a clear opportunity to study the mineralogy and composition of: (1) an ancient fluvio-lacustrine sedimentary deposit, (2) aqueously altered early martian material, and (3) a regional carbonate unit unique to the region around the Isidis impact basin. Therefore, it is critical to understand the mineralogy of the rover-accessible units within Jezero crater using presently available data to maximize the potential for landed science and highlight the science questions that may be addressed by such a mission. However, following the initial identification of phyllosilicate and carbonate minerals within the Jezero crater basin (Ehlmann et al., 2008b, 2008a), most workers have primarily focused on regional views of the mineralogy of Nili Fossae (Ehlmann et al., 2009) and the Jezero crater watershed (Goudge et al., 2015). There has yet to be a detailed study of the mineralogy of alteration mineral-bearing units contained within the crater itself and how they correlate to spectrally similar regional units.

In this study, we present remote observations of the mineralogy of phyllosilicate- and carbonate-bearing units contained within Jezero crater that may be accessible to a future landed mission. We identify and map a range of olivine-bearing lithologies that are present in the Nili Fossae and Jezero crater region, which we hypothesize can be discriminated by olivine Mg-content. We demonstrate that these lithologies correspond to previously mapped geomorphic units (Goudge et al., 2015). We show that, intriguingly, the carbonate occurrences are always hosted in a relatively Fe-rich olivine-bearing lithology, and we discuss the implications for the formation and later alteration of the rocks exposed at Jezero crater.

### 1.1 Jezero Crater Delta and Mineralogy

The two delta deposits within Jezero contain Fe/Mg-phyllosilicate and carbonate in varying proportions, with the northern fan dominated by carbonate and the western fan dominated by phyllosilicate (Goudge et al., 2015). The provenance of the phyllosilicate and carbonate within the Jezero crater deltas can be traced to mineralogically similar protolith units within the watershed, which provides strong evidence that the alteration minerals were primarily emplaced by fluvial transport (Goudge et al., 2015). These deposits therefore have bulk compositions that integrate heavily altered martian crust of the Nili Fossae region (Ehlmann et al., 2009; Goudge et al., 2015; Mustard et al., 2009), and offer an opportunity to examine a diverse array of alteration minerals. Furthermore, Jezero crater contains large exposures of olivine- and carbonate-bearing units on the floor of the crater, underlying the deltaic deposits, and draping the interior rim of the crater (Ehlmann et al., 2008b, 2009; Goudge et al., 2015). These deposits were interpreted by Goudge et al. (Goudge et al., 2015) to represent exposures of the regional olivine-carbonate-bearing unit observed elsewhere in Nili Fossae (Ehlmann et al., 2008b, 2009) and are similar to carbonate-bearing geomorphic units mapped recently in a study of the neighboring northeast Syrtis Major region (Bramble et al., 2017).

## 2 Methods

We have analyzed data from the Compact Reconnaissance Imaging Spectrometer for Mars (CRISM) (Murchie et al., 2007) instrument covering the Jezero crater and watershed for the presence of primary and alteration minerals exposed at the surface. We used the CRISM image BD1300 summary parameter (Viviano-Beck et al., 2014) to produce an olivine discriminating map of the Jezero crater watershed. A detailed list of CRISM observations and mineral identifications is listed in the Supplementary Material (SM) accompanying this paper.

The BD1300 parameter measures the band depth at 1.32 μm relative to a continuum defined by the reflectances at 1.08 and 1.75 μm. The parameter is sensitive to the broad olivine 1 μm absorption band, which has a short wavelength shoulder at 0.6-0.7 μm, and a long wavelength shoulder at 1.7-1.8 μm. The position of the 1 μm olivine band is sensitive to the Mg vs. Fe content, where more Fe-rich olivine displays a longer 1 μm band position (King and Ridley, 1987). Because the BD1300 parameter is sensitive to the depth of the 1 μm band at a longer wavelength position, it will display higher values for Fe-rich olivine than Mg-rich olivine. The chief advantage of this approach to mapping olivine mineralogy is that its simplicity allows it to be applied to any CRISM image, including those obtained with a subset of wavelengths in the instrument's "multispectral mapping" mode (see SM for further detail).

## 3 Results

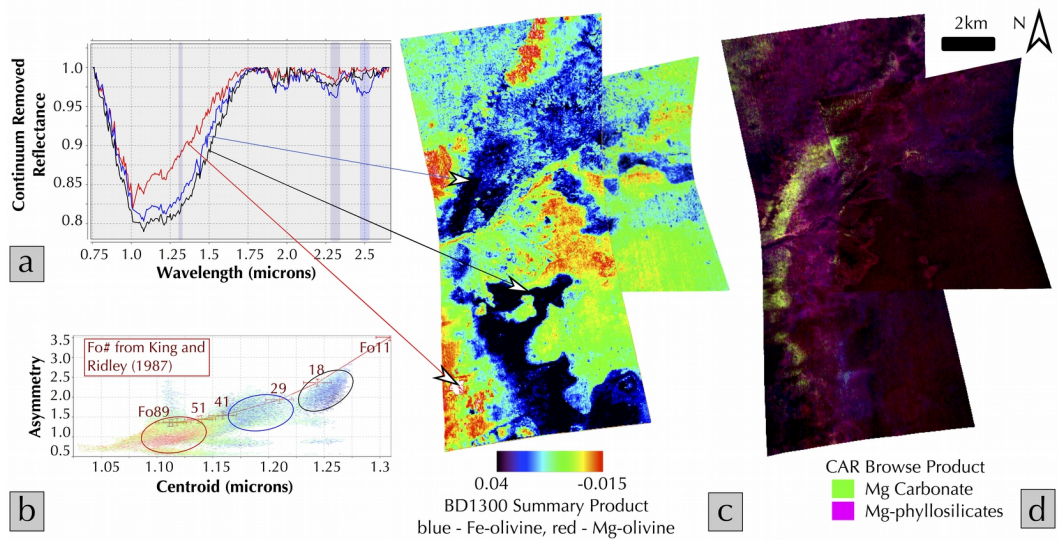

**Figure 1. a)** Example continuum removed spectra from arrowed locations in image HRL40FF. The 1.3, 2.3 and 2.5 μm regions are indicated by shaded vertical regions. **b)** Plot of asymmetry versus centroid position for the 1 μm absorption band of library olivine spectra with known Fo/Mg #, overlain on the points for HRL40FF color-coded for the BD1300 parameter used in the central image. **c)** BD1300 map of Jezero western delta covered by CRISM images HRL40FF and FR47A3. Arrows show regions where example spectra were obtained. **d)** CRISM CAR standard browse products (Viviano-Beck et al., 2014) of Jezero delta, showing regions where carbonate is present, due to the presence of a 2.3 accompanied by a 2.5 μm band.

### 3.1 Olivine Composition

Figure 1a shows continuum removed spectra from a CRISM image over the Jezero delta at three representative points within the crater (locations shown by arrows in Figure 1c). The continuum removed spectra show a distinct shift in the wide olivine 1 μm band, which we interpret as indicative of a range of Mg-content (King and Ridley, 1987). The black spectrum represents the units mapped in black, which contains the most Fe-rich olivine unit. The blue spectrum represents an example of an olivine-carbonate spectrum that is exposed on the west of the crater, and also displays carbonate bands at 2.3 and 2.5 μm. The red spectrum corresponds to a lithology that contains the most Mg-rich olivine, mapped in red in the BD1300 map.

Figure 1b shows a plot of the asymmetry versus the centroid for the olivine 1 μm band, both fit by an asymmetric Gaussian curve (Brown, 2006; Brown et al., 2010) (see SM for further discussion of these metrics) for the HRL40FF data shown in Figure 1c. Also shown in this plot are the centroids and asymmetries of USGS library olivine spectra of known forsterite-fayalite composition (Clark et al., 2007). Using this plot, we can estimate (using the extent of clusters shown in color-matched ellipses) that our typical three lithologies have Fo# ranges of 15-25, 25-40 and 50-90, broadly corresponding to the black, blue and red/white units in the BD1300 image, respectively. Figure 1d shows the carbonate browse (CAR) image which highlights Mg-carbonate, Mg/Fe-phyllosilicates and hydrated minerals (Viviano-Beck et al., 2014) over the same CRISM images, which demonstrates that the largest carbonate deposits in the crater are associated with the unit that is mapped in blue in the BD1300 parameter map. The highest Mg# olivine-bearing lithology, mapped in red-white, also outcrops on the crater rim and the edges of the western delta deposit.

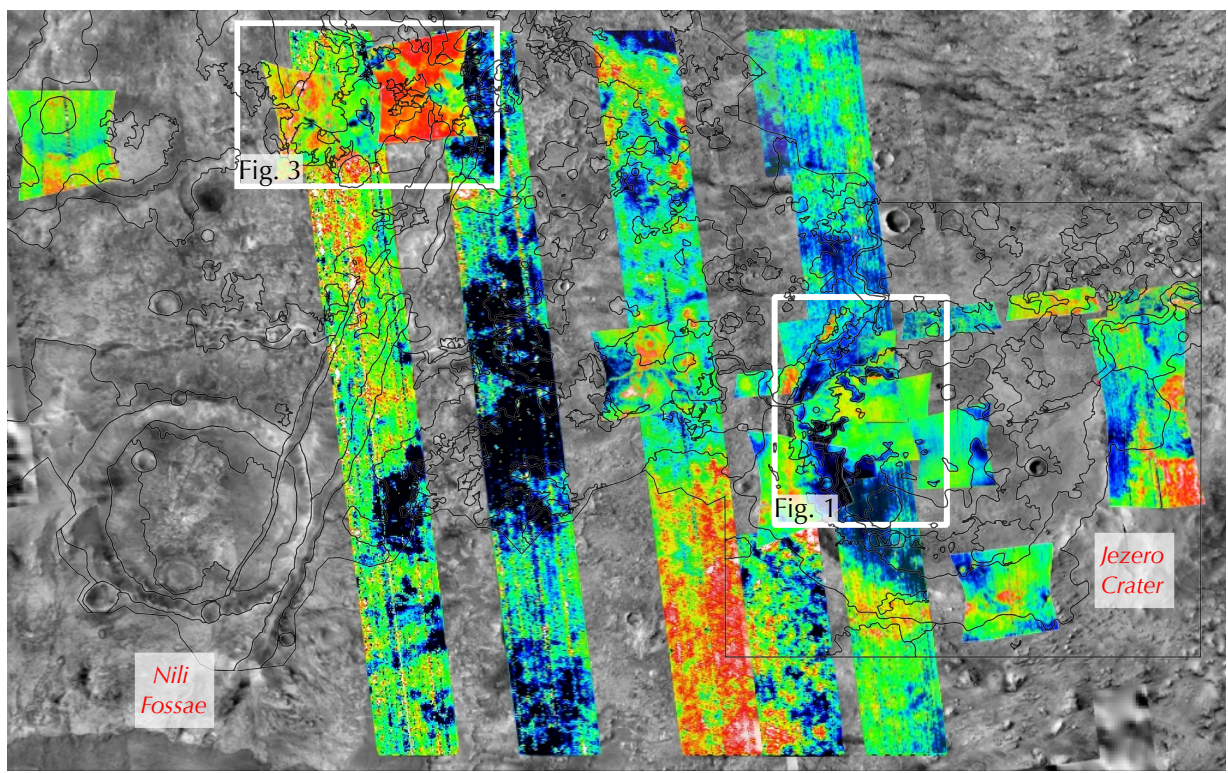

**Figure 2.** CRISM BD1300 map of the Jezero crater and delta watershed, overlying a mosaic of CTX images of the Jezero watershed region. Thin black lines show the boundaries of major geomorphic units mapped in Goudge et al. (2015). The small images are CRISM hyperspectral targeted observations and the long strips are multispectral mapping data.

Figure 2 represents a comparison of the BD1300 map of the Jezero crater and delta watershed with the existing geomorphic map produced in ref. 16. Figure 2 demonstrates a strong correlation between the boundaries of the geomorphic units of this previous mapping study (Goudge et al., 2015) and variations in the CRISM BD1300 map. This suggests that the three olivine-bearing lithologies that we identified here are contained within distinct and recognizable geomorphic units in the watershed, as well as within surrounding region, suggesting a consistent regional geologic alteration framework throughout the eastern Nili Fossae region, as proposed by Viviano et al. (2013).

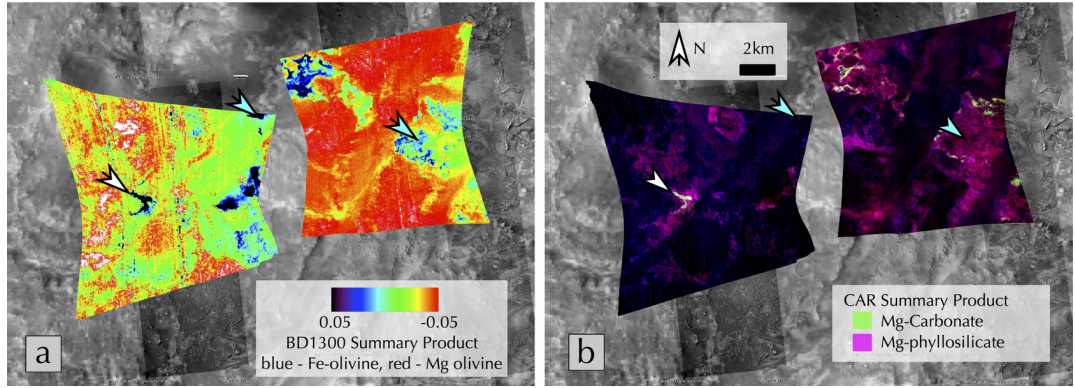

**Figure 3. a)** Olivine mineralogy of Nili Fossae as mapped by CRISM images FRT23370 and 97E2 overlying a CTX basemap with HiRISE images where available. **b)** Carbonate standard browse product (CAR) showing region where carbonate is mapped. Carbonate is in green and 2.3 band material (Mg-phyllosilicate) is in purple. Note that carbonates are only associated with the blue-black olivine unit, and the blue olivine unit is only partially carbonatized (white arrows indicate where carbonate has formed, light blue arrows show where it has not formed).

### 3.2 Partial Carbonization

Figure 3 presents a BD1300 map of a region of the Nili Fossae (location outlined in Figure 2). Figure 3 demonstrates a key property of the carbonation process of the most Fe-rich olivine-bearing lithology: the carbonate-bearing regions of this lithology are spatially variable. The white arrows on the figure indicate that when the carbonate is present, it is located within an Fe-olivine lithology. The light blue arrows on the left and right show regions of the Fe-olivine-bearing lithology (shown in blue on the left) that have not been carbonatized (since they are not green on the right). This pattern is also visible in the Jezero crater itself, where the Fe-olivine-lithology is variably carbonatized (Figure 1c and d).

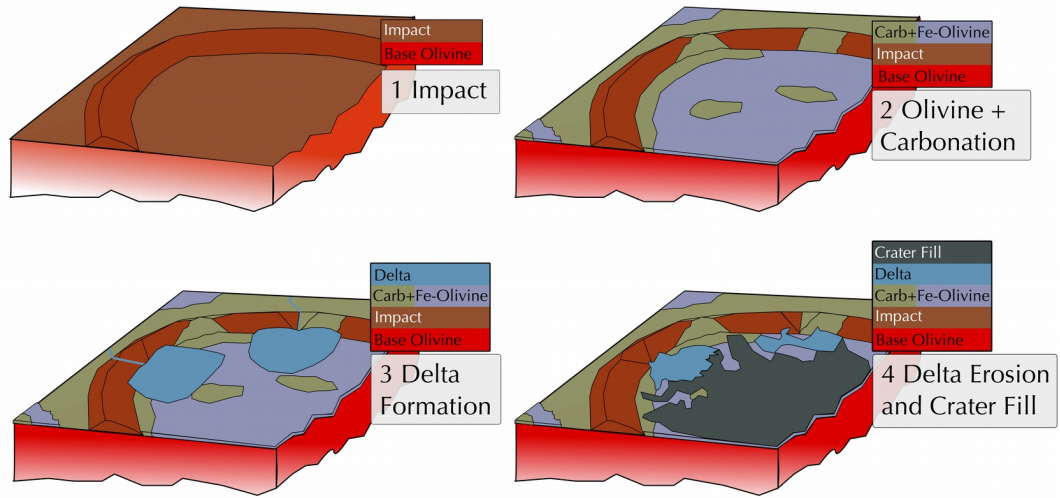

**Figure 4.** Framework Formation history for Jezero crater carbonates and olivine units that are relevant to our findings. Step 1 is the impact that formed Jezero, Step 2 is the emplacement of the Fe-rich olivine-bearing lithology and the variable carbonatization of this lithology, Step 3 is the emplacement of the two deltas and Step 4 is erosion of the deltas and emplacement of crater infill, leading to the crater as we see it today.

## 4 Discussion

An idealised geological sequence of events at Jezero crater is shown in Figure 4. This outlines our current interpretation of the timing of major geological processes that have shaped the region summarised in four steps. Step 1 in Figure 4 is the impact that formed Jezero Crater, which postdates the Nili Fossae formation. Step 2 is emplacement of the Fe-olivine bearing lithology that we have mapped in Figures 1-3, and partial carbonatisation of this lithology. Our finding that there are no exposures of the Fe-olivine bearing lithology in the crater rims (Figures 1 and 2) is strong evidence that this unit was not present when the impact occurred, and post-dates it, and this is consistent with previous stratigraphic interpretations of this unit (Goudge et al., 2015). Step 3 is the formation of fluvial valleys, filling of the crater with water and the emplacement of deltaic deposits, prior to or during the Late Noachian-Early Hesperian valley network forming period (Fassett and Head, 2008; Irwin et al., 2005). Step 4 is the erosion of the delta deposits to their current degraded state and infill of the crater by the floor resurfacing unit (Goudge et al., 2015; Schon et al., 2012). The latter two steps must post-date the carbonate-formation, as the northern and western delta appear to contain detrital carbonate transported to the basin from the watershed (Goudge et al., 2015).

If future studies are able to determine or better constrain the formation and alteration scenarios at Jezero, then we will be able to address a number of outstanding science questions that arise from the identification of the olivine-bearing lithologies reported in this study. We now highlight three such key science questions that have arisen from the results of this study of the olivine-carbonate lithology.

### 4.1. Emplacement of Mg-olivine and Fe-olivine lithologies

The first question posed by our findings is perhaps the most obvious: why is the Fe-rich olivine-bearing lithology preferentially altered to carbonate, while the more Mg-rich olivine-bearing lithologies are not associated with carbonate? Typical terrestrial studies of ultramafic serpentinisation examine the Mg-$SiO_2$-$H_2O$ sequence (Moody, 1976), however some serpentinisation of Fe-bearing olivine has been recently investigated (Klein et al., 2013). Skok et al. (Skok et al., 2012) used the Modified Gaussian Model approach to calculate the Mg# of CRISM observations of Martian crater central peaks, in order to access the most primitive Martian crust samples, and found a range of Fo#5-60 in the craters they studied. This is similar to our estimated Fo# for the Fe-rich olivine-bearing lithology we have mapped here. The high Fo# for the older exposures of olivine-bearing lithology was not reported by Skok et al., indicating these Mg-rich olivine exposures may be a key, as yet unrecognised, early Martian olivine composition. In situ analysis of these olivine lithologies to discover whether there are further geochemical or mineralogical differences between them is likely to have implications for the formation of the entire mafic surface of Mars.

### 4.2. Partial Carbonatisation

The second question arising from our study is: why is the Fe-olivine-bearing lithology only partially carbonatized? One hypothesis that addresses the restriction of carbonatization to only parts of the Fe-bearing olivine lithology (and none of the underlying or overlying units) is that the alteration was driven by relatively weak temperature and pressure profiles, perhaps driven by heat associated with emplacement of the Fe-bearing olivine unit. Another hypothesis is that spatial alteration variations may be due to permeability in fracture structures through which hydrothermal fluids could circulate.

### 4.3. Associated Mg-phyllosilicate minerals

The third question our study raises is: what is the relationship of carbonates to associated Mg-phyllosilicate signatures? The carbonate signatures in Figure 3 are accompanied by a 2.38 μm absorption band that is indicative of Mg-phyllosilicates (Ehlmann et al., 2009) of uncertain mineralogy (Brown et al., 2010; Viviano et al., 2013). *In situ* investigation of the mineralogy associated with the carbonate layer, including the identification of specific Mg-phyllosilicates such as chlorite (Brown et al., 2005; Viviano et al., 2013), saponite (Ehlmann et al., 2009), serpentine (Brown et al., 2005; Ehlmann et al., 2009) or talc (Brown et al., 2010; Viviano et al., 2013), will help reveal the alteration conditions and temperature and pressure conditions that accompanied the carbonization, and may determine whether it was a serpentinization, sedimentary or leaching process. The formation conditions are crucial because the presence of talc-carbonate resulting from the carbonation of serpentine has been examined in Earth analogs in terrestrial greenstone belts such as the Pilbara in Western Australia (Brown et al., 2004, 2005, 2006), where talc-bearing komatiite cumulate units of the Dresser Formation overlie the siliceous, stromatolite-bearing Strelley Pool Chert unit (Van Kranendonk et al., 2008). An in situ investigation of the Mg-phyllosilicate mineralogy is therefore a critical task in understanding the astrobiological potential of the carbonate deposits at Nili Fossae.

Table 1 presents a summary of potential formation scenarios we are presently aware of that may have played a role in the emplacement and alteration of the olivine-carbonate lithology in the Jezero crater region. Each of the formation and alteration scenarios in Table 1 present a plausible formation mechanism for the emplacement of the carbonate at Nili Fossae, and potentially the accompanying olivine and associated Mg-phyllosilicates. We have made a qualitative assessment of the ability of each scenario to address the findings of this study, based on current descriptions in the literature. At this time we do not have the required data to conclude which scenario is correct, and/or whether multiple scenarios were at play. However, in the last column of the table, we have provided testable hypotheses and/or observables for each scenario that could be used to evaluate each mechanism with future *in situ* exploration.

| Scenario | 1. Olivine Placement | 2. Partial carbonatisation | 3. Associated Mg-phyllo | In situ testable hypothesis or observable |
|---|---|---|---|---|
| Superposition of a volcanic succession on early Martian crust, with accompanying deuteric alteration; serpentinization reactions driven by heat of volcanic/komatiite flows (Brown et al., 2010; Viviano et al., 2013) | √√√ | √√√ | √√√ | Lava flows in stratigraphic section |
| Impact-driven hydrothermal activity by a melt sheet; serpentinization reactions driven by hydrothermal activity from direct heat of impact or impact melt (Osinski et al., 2013) | √√ | √√√ | √√ | Superposed impact melt sheet in stratigraphic section |
| Subsurface alteration under thicker $CO_2$ atmosphere; serpentinization reactions driven by diagenesis and upper crustal hydrothermal processes (van Berk and Fu, 2011) | √ | √√√ | √√√ | Indicators of pedogenic alteration in carbonate and layering as predicted in ref. 33. |
| Hydrothermal alteration in thermal springs environment (Walter and Des Marais, 1993) or alteration of volcanic tephra by ephemeral waters (Ruff et al., 2014) | √ | √√√ | √√ | Mineralogical and physical evidence of tephra-like deposits |
| Cold ophiolite-hosted serpentinization, as in the terrestrial analogs in California (Campbell et al., 2002; Schulte et al., 2006) or the Oman ophiolite (Paukert et al., 2012) | √ | √√ | √ | Low temperature serpentinization minerals (See discussion) |
| Low temperature leaching, as in terrestrial analog of Antarctic carbonate rinds (Doran et al., 1998; Salvatore et al., 2013) and Mojave desert carbonate rinds (Bishop et al., 2011) | √ | √√ | √ | Carbonate in subsurface rinds |
| Precipitation of carbonate directly into a marine basin, this includes scenarios of an early Martian ocean preserved at Nili Fossae (Russell et al., 2014) | | √ | √ | Exhalative/smoker structures or stromatolite morphology |
| Hydrothermal formation of clays from an olivine bearing Martian basalt under a thick $CO_2$ atmosphere (Dehouck et al., 2014) or under a high pressure and temperature atmosphere (Cannon et al., 2017) | √√√ | √ | √ | High temperature or pressure mineral phases (see Discussion) |
| Transport and deposition of olivine/carbonate sediment in a large aeolian dune field, such as the lower unit of the Burns Formation (Grotzinger et al., 2005) | √√ | √ | √ | Aeolian sedimentary features in the carbonate |

**Table 1.** Possible Formation scenarios, science questions arising from this study and testable predictions for Jezero olivine-carbonate sequence. No ticks indicates the scenario cannot address this question. One tick indicates the scenario might account for this question, two ticks indicates it partially deals with the question, and three ticks means it specifically addresses with this question.

## 5 Conclusions

We have presented evidence that the olivine-carbonate lithology mapped at Jezero crater constitutes a regional unit that represents a key sequence in the early history of Mars. Several questions regarding the Martian interior can be addressed by these newly recognized associations. If the olivine lithologies are found to be in regional volcanic eruptive layers traceable to a single subsurface heat source, the Mg to Fe variation we have reported may indicate a sequence of Mg-olivine in the base, followed by a hiatus, then a more Fe-rich sequence, as expected for a depleted or cooling mantle source through time (Condie et al., 2016). If the origin of the olivine lithologies is volcanic, the mineralogy of the landing site could be used to constrain early martian mantle potential temperatures (Putirka, 2016) at 3.9 Ga (the estimated formation age of Isidis Basin (Werner et al., 2011)) and further our understanding of how the martian interior has cooled through time, via plumes in a stagnant lid regime (Stevenson, 2003). Given the importance of tectonic regime for the evolution of life on Earth (Nisbet and Sleep, 2001), the rate of interior cooling is important. If life were discovered, and the cooling regime verified, it would suggest that a plate tectonic driven water cycle is not necessary, at least for simple life forms to evolve (although a tectonic-mediated water cycle might still be needed to maintain surficial life for billions of years). On the other hand, if the stagnant lid cooling regime is verified and life is absent, this may support the proposal that even very simple life forms require some form of plate tectonics to gain their start.

The results of this study suggest that *in situ* exploration of the olivine-carbonate-bearing lithologies at Jezero crater would deepen our understanding of Noachian Mars, while at the same time teaching us about the styles of tectonics and astrobiological potential of small exoplanets outside our solar system (such as Kepler 138b (Jontof-Hutter et al., 2015)), that are now being found to be more plentiful in our galaxy than Earth-sized planets.

**Acknowledgments, Samples, and Data**

The authors declare no financial conflicts. All CRISM data used can be obtained from the Planetary Data System (PDS). AJB conducted this study with the support of the NASA Astrobiology Institute, (Grant# NNX15BB01A) and NASA MDAP (Grant# NNX16AJ48G). TAG acknowledges support from the CRISM team through a subcontract from the Johns Hopkins University Applied Physics Lab. We would like to thank Keith Putirka for discussions on early Martian core conditions, and Mike Russell for constant inspiration. We would also like to thank Scott Murchie for reviewing a draft of this paper and his CRISM team at JHUAPL for their untiring work to produce a fantastic dataset.